# Characteristics of Minimal Effective Programming Systems[*]

Samuel E. Moelius III
IDA Center for Computing Sciences
17100 Science Drive, Bowie, MD 20715-4300
semoeli@super.org

August 20, 2018

**Abstract.** The Rogers semilattice of effective programming systems (epses) is the collection of all effective numberings of the partial computable functions ordered such that $\theta \leq \psi$ whenever $\theta$-programs can be algorithmically translated into $\psi$-programs. Herein, it is shown that an eps $\psi$ is minimal in this ordering if and only if, for each translation function $t$ into $\psi$, there exists a computably enumerable equivalence relation (ceer) $R$ such that (i) $R$ is a subrelation of $\psi$'s program equivalence relation, and (ii) $R$ equates each $\psi$-program to some program in the range of $t$. It is also shown that there exists a minimal eps for which no *single* such $R$ does the work for all such $t$. In fact, there exists a minimal eps $\psi$ such that, for each ceer $R$, either $R$ contradicts $\psi$'s program equivalence relation, or there exists a translation function $t$ into $\psi$ such that the range of $t$ *fails* to intersect *infinitely many* of $R$'s equivalence classes.

## 1 Introduction

Let $\mathbb{N}$ be the set of natural numbers, i.e., $\{0, 1, 2, ...\}$. An effective programming systems (eps) is a partial computable function $\lambda p, x . \psi_p(x)$ mapping $\mathbb{N}^2$ to $\mathbb{N}$, and having the following property. For each partial computable function $\zeta$ mapping $\mathbb{N}$ to $\mathbb{N}$, there exists a $p$ such that $\psi_p = \zeta$. Effective programming systems abstract the notion of *programming language* in the following sense. One can think of $p$ as a *program*, and of $\psi_p$ as the partial computable function denoted by $p$ within some programming language corresponding to $\psi$.

Rogers [Rog58] introduced the following ordering on epses. For epses $\theta$ and $\psi$, $\theta \leq \psi$ iff there exists a computable function $t : \mathbb{N} \to \mathbb{N}$ such that, for each $p$, $\theta_p = \psi_{t(p)}$. Intuitively, $\theta \leq \psi$ whenever $\theta$-programs can be algorithmically translated into $\psi$-programs. Moreover, an eps $\psi$ is *minimal* in this ordering iff having the ability to algorithmically translate $\theta$-programs into $\psi$-programs implies having the ability to algorithmically translate $\psi$-programs into $\theta$-programs, for each eps $\theta$.

Arguably, the most well studied collection of minimal epses is that of the Friedberg numberings [Fri58, Kum90]. Recall that a Friedberg numbering is an eps that is 1-1, i.e., for each $p$ and $q$, $\psi_p = \psi_q$ implies $p = q$. Examples of works that make use of this concept include [Lav77, MWY78, Ric81, FKW82, Sch82, Roy87, Kum89, Spr90, GYY93, HK94, JST11].

In [PE64], Pour-El asked whether every minimal eps is equivalent to some Friedberg numbering. Ershov [Ers68, §5] showed that there exists a minimal effective numbering of the *computably enumerable sets* that is not equivalent to any 1-1 numbering. Shortly thereafter, his student, Khutoretskii, established the analogous result for the partial computable functions, thereby answering Pour-El's question.

**Theorem 1 (Khutoretskii [Khu69a, Ex. 1 and Cor. 4]).** *There exists a minimal eps that is not equivalent to any Friedberg numbering.*

For the purposes of this paper, Theorem 1 is best viewed through the following folklore theorem. (For completeness we give a proof of this result.)

**Theorem 2 (Folklore).** *For each eps $\psi$, $\psi$ is equivalent to a Friedberg numbering iff $\psi$'s program equivalence relation is computable.*

---

[*] This is an expanded version of [Moe12].

*Proof.* Let $\psi$ be given.

($\Rightarrow$) Suppose that $\psi$ is equivalent to a Friedberg numbering $\eta$, and that $t : \mathbb{N} \to \mathbb{N}$ witnesses $\psi \leq \eta$. Then, clearly, for each $p$ and $q$,
$$\psi_p = \psi_q \Leftrightarrow \eta_{t(p)} = \eta_{t(q)} \Leftrightarrow t(p) = t(q). \tag{1}$$
Thus, since $\lambda p, q . [t(p) = t(q)]$ is computable, $\psi$'s program equivalence relation is computable.

($\Leftarrow$) Suppose that $\psi$'s program equivalence relation is computable. Let $M$ be the set of *minimal programs* in $\psi$, i.e., $M = \{m_0, m_1, ...\}$ where, for each $i$, $m_i$ is *least* such that
$$\psi_{m_i} \notin \{\psi_{m_0}, ..., \psi_{m_{i-1}}\}. \tag{2}$$
Note that, since $\psi$'s program equivalence relation is computable, $M$ is computable. Let $\eta$ be such that, for each $i$,
$$\eta_i = \psi_{m_i}. \tag{3}$$
Using the fact the $M$ is computable, it is straightforward to verify that $\eta$ is a Friedberg numbering, and that $\psi \equiv \eta$. $\square$ (**Theorem 2**)

In light of Theorem 2, Theorem 1 may be restated as: there exists a minimal eps whose program equivalence relation is *not* computable. On the other hand, as noted in the proof of Theorem 1, the constructed eps's program equivalence relation is computably enumerable. (In particular, exactly one such equivalence class is a simple set [Rog67, §8.1], and all others a singletons.) Thus, one has the following.

**Theorem 3 (Khutoretskii, corollary of Thm. 2 and proof of Thm. 1).** There exists an eps whose program equivalence relation is computably enumerable, but *not* computable.

Subsequent to the above, Khutoretskii showed the following.

**Theorem 4 (Khutoretskii, corollary of [Khu69b, Thm. 1]).** There exists a minimal eps whose program equivalence relation is *not* computably enumerable.

Clearly, Theorems 3 and 4 can be viewed as a sharpening of Theorem 1. Herein, we sharpen Khutoretskii's results even further.

To facilitate the statement of our results, we first give a few definitions. Suppose that $\psi$ is an eps. For each $t : \mathbb{N} \to \mathbb{N}$, we say that $t$ is a *translation function into* $\psi$ iff there exists an eps $\theta$ such that $t$ witnesses $\theta \leq \psi$. The following definition is equivalent. For each $t : \mathbb{N} \to \mathbb{N}$, $t$ is a translation function into $\psi$ iff $t$ is computable and the partial function $\lambda p, x . \psi_{t(p)}(x)$ is an eps.

**Definition 5.** Suppose that $\psi$ is an eps, and that $t$ is a translation function into $\psi$. Then, for each equivalence relation $R$, (a) and (b) below.

(a) $R$ *strongly ties* $t$ *into* $\psi$ iff $R$ satisfies (i) and (ii) just below.[1]
   (i) $R$ is a subrelation of $\psi$'s program equivalence relation.
   (ii) The range of $t$ intersects each of $R$'s equivalence classes.
(b) $R$ *weakly ties* $t$ *into* $\psi$ iff $R$ satisfies (i) just above and (ii*) just below.[2]
   (ii*) The range of $t$ intersects all but finitely many of $R$'s equivalence classes.

Thus, if equivalence relation $R$ strongly ties translation function $t$ into eps $\psi$, then $R$ equates each $\psi$-program to some program in the range of $t$. If $R$ merely weakly ties $t$ into $\psi$, then there may be infinitely many $\psi$-programs that $R$ does *not* equate to any program in the range of $t$. However, those infinitely many such $\psi$-programs will form only finitely many equivalence classes.

Our first main result is that the minimal epses may be *characterized* as follows.

**Theorem 6.** For each eps $\psi$, (a)-(c) below are equivalent.

(a) $\psi$ is minimal.

---
[1] In some places, we omit the phrase "into $\psi$" when it is clear from context.
[2] See footnote 1.

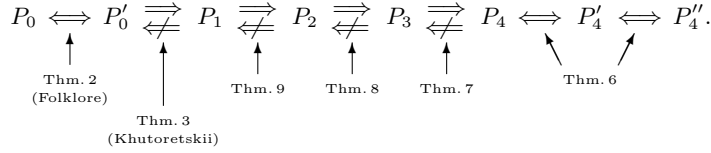

- $P_0(\psi) \Leftrightarrow \psi$ is equivalent to a Friedberg numbering.
- $P'_0(\psi) \Leftrightarrow \psi$'s program equivalence relation is computable.
- $P_1(\psi) \Leftrightarrow \psi$'s program equivalence relation is computably enumerable.
- $P_2(\psi) \Leftrightarrow$ there exists a ceer $R$ that strongly ties each translation function into $\psi$.
- $P_3(\psi) \Leftrightarrow$ there exists a ceer $R$ that weakly ties each translation function into $\psi$.
- $P_4(\psi) \Leftrightarrow$ for each translation function $t$ into $\psi$, there exists a ceer that strongly ties $t$ into $\psi$.
- $P'_4(\psi) \Leftrightarrow$ for each translation function $t$ into $\psi$, there exists a ceer that weakly ties $t$ into $\psi$.
- $P''_4(\psi) \Leftrightarrow \psi$ is minimal.

**Fig. 1.** A summary of the results mentioned in Section 1. In addition to the above: Mal'cev [Mal65, Mal71] showed that $P_1 \Rightarrow P''_4$, and Khutoretskii [Khu69b] showed that $P_1 \not\Leftarrow P''_4$ (see Theorem 4).

(b) For each translation function $t$ into $\psi$, there exists a computably enumerable equivalence relation (ceer)[3] that strongly ties $t$ into $\psi$.
(c) For each translation function $t$ into $\psi$, there exists a ceer that weakly ties $t$ into $\psi$.

Note that Theorem 4 is about a *single* equivalence relation, i.e., the program equivalence relation of a certain eps, whereas Theorem 6 is about one equivalence relation *per* translation function into any given eps. Thus, one might ask: if $\psi$ is a minimal eps, then might there always exist a *single* ceer that strongly ties each translation function into $\psi$? The answer, as it turns out, is *no*. In fact, as Theorem 7 below states, there need not even exist a single ceer that *weakly* ties each translation function into $\psi$.

**Theorem 7.** There exists an eps $\psi$ satisfying (a) and (b) below.

(a) $\psi$ is minimal.
(b) For each ceer $R$, there exists a translation function $t$ into $\psi$ such that $R$ does *not* weakly tie $t$ into $\psi$.

Continuing with this line of thought, one finds that the strong and weak notions of Definition 5 separate when one considers single equivalence relations.

**Theorem 8.** There exists an eps $\psi$ and a ceer $R$ satisfying (a) and (b) below.

(a) For each translation function $t$ into $\psi$, $R$ weakly ties $t$ into $\psi$.
(b) For each ceer $R'$, there exists a translation function $t$ into $\psi$ such that $R'$ does *not* strongly tie $t$ into $\psi$.

Clearly, if $\psi$ is an eps, and $\psi$'s program equivalence relation is computably enumerable, then there exists a single ceer $R$ that strongly ties each translation function into $\psi$, i.e., $R$ is $\psi$'s program equivalence relation. Thus, one might ask: does the converse hold? Theorem 9, just below, establishes that it does *not*.

**Theorem 9.** There exists an eps $\psi$ and a ceer $R$ satisfying (a) and (b) below.

(a) For each translation function $t$ into $\psi$, $R$ strongly ties $t$ into $\psi$.
(b) $\psi$'s program equivalence relation is *not* computably enumerable.

Figure 1 summarizes the results mentioned in this section. The remainder of this paper is organized as follows. Section 2 covers preliminaries. Section 3 gives complete proofs of Theorems 6 through 9.

---

[3] We pronounce ceer like the first syllable of "series". Computably enumerable equivalence relations are of interest in their own right. Gao and Gerdes [GG01] give an excellent survey.

## 2 Preliminaries

Computability-theoretic concepts not covered below are treated in [Rog67].

Lowercase math-italic letters (e.g., $i$, $p$, $x$), with or without decorations, range over elements of $\mathbb{N}$, unless stated otherwise. Uppercase math-italic letters (e.g., $I$, $P$, $X$), with or without decorations, range over subsets of $\mathbb{N}$, unless stated otherwise. For each *non*-empty $X$, $\min X$ denotes the minimum element of $X$. $\min \emptyset \stackrel{\text{def}}{=} \infty$. For each *non*-empty, finite $X$, $\max X$ denotes the maximum element of $X$. $\max \emptyset \stackrel{\text{def}}{=} -1$. *Fin* denotes the collection of all finite subsets of $\mathbb{N}$.

$\langle \cdot, \cdot \rangle$ denotes any fixed pairing function, i.e., a 1-1, onto, computable function of type $\mathbb{N}^2 \to \mathbb{N}$ [Rog67, page 64]. For each $x$, $y$, and $z$, $\langle x, y, z \rangle \stackrel{\text{def}}{=} \langle x, \langle y, z \rangle \rangle$. For each $X$ and $Y$, $X \times Y \stackrel{\text{def}}{=} \{\langle x, y \rangle \mid x \in X \land y \in Y\}$.

Every partial function considered herein maps $\mathbb{N}$ to $\mathbb{N}$, unless stated otherwise. For each partial function $\zeta$, and each $x$, $\zeta(x)\downarrow$ denotes that $\zeta(x)$ converges; whereas, $\zeta(x)\uparrow$ denotes that $\zeta(x)$ diverges. We use $\uparrow$ to denote the value of a divergent computation. For the sake of some subsequent proofs, it is convenient to have the following notation. For each $i$ and $n$,

$$i^{<n} \stackrel{\text{def}}{=} \lambda x \,.\, \begin{cases} i, & \text{if } x < n; \\ \uparrow, & \text{otherwise.} \end{cases} \tag{4}$$

Thus, $i^{<n}$ is the partial function that maps each value less than $n$ to $i$, and that diverges everywhere else. For each partial function $\zeta$, $\mathrm{rng}(\zeta)$ denotes the range of $\zeta$, i.e., $\mathrm{rng}(\zeta) \stackrel{\text{def}}{=} \{y \mid (\exists x)[\zeta(x) = y]\}$. *PartComp* denotes the set of all partial computable functions (mapping $\mathbb{N}$ to $\mathbb{N}$).

$\varphi$ denotes any fixed acceptable (i.e., maximal) eps [Rog58, Rog67, MWY78, Ric81, Roy87]. For each $p$, $W_p \stackrel{\text{def}}{=} \{x \mid \varphi_p(x)\downarrow\}$. For each $p$ and $s$, the following.

$$\varphi_p^s \stackrel{\text{def}}{=} \lambda x \,.\, \begin{cases} \varphi_p(x), & \text{if } x < s \text{ and } \varphi_p(x) \text{ converges in fewer than } s \text{ steps}; \\ \uparrow, & \text{otherwise.} \end{cases} \tag{5}$$

$$W_p^s \stackrel{\text{def}}{=} \{x \mid \varphi_p^s(x)\downarrow\}. \tag{6}$$

For each eps $\psi$, $\mathrm{Equiv}(\psi)$ denotes $\psi$'s program equivalence relation, i.e.,

$$\mathrm{Equiv}(\psi) \stackrel{\text{def}}{=} \{\langle p, q \rangle \mid \psi_p = \psi_q\}. \tag{7}$$

For each equivalence relation $R$, *Classes*$(R)$ denotes the set of $R$'s equivalence classes, i.e., *Classes*$(R)$ is the set of exactly those $E$ satisfying (a)-(c) below.

(a) $E \neq \emptyset$.
(b) $(\forall p, q \in E)[\langle p, q \rangle \in R]$.
(c) $(\forall p \in E)(\forall q \notin E)[\langle p, q \rangle \notin R]$.

## 3 Results

This section recounts our main results (Theorem 6 through 9), and gives their complete proofs.

Our first main result is that the minimal epses may be *characterized* as per Theorem 6, restated just below. Recall from Definition 5 that if equivalence relation $R$ strongly ties translation function $t$ into eps $\psi$, then (i) $R$ is a subrelation of $\psi$'s program equivalence relation, and (ii) the range of $t$ intersects each of $R$'s equivalence classes. On the other hand, if $R$ merely weakly ties $t$ into $\psi$, then the range of $t$ need only intersect all but finitely many of $R$'s equivalence classes.

**Theorem 6.** For each eps $\psi$, (a)-(c) below are equivalent.

(a) $\psi$ is minimal.
(b) For each translation function $t$ into $\psi$, there exists a ceer that strongly ties $t$ into $\psi$.
(c) For each translation function $t$ into $\psi$, there exists a ceer that weakly ties $t$ into $\psi$.

*Proof.* Let $\psi$ be given.

(a) $\Rightarrow$ (b): Suppose that $\psi$ is minimal. Let $t$ be any translation function into $\psi$, and let $\theta$ be such that $t$ witnesses $\theta \leq \psi$. Since $\psi$ is minimal, there exists a $t' : \mathbb{N} \to \mathbb{N}$ witnessing $\psi \leq \theta$. Let $R$ be the reflexive, symmetric, transitive closure of

$$\{\langle p, (t \circ t')(p)\rangle \mid p \in \mathbb{N}\}. \tag{8}$$

Clearly, $R$ is a ceer and $R \subseteq \mathrm{Equiv}(\psi)$. It remains to show that, for each $E \in \mathit{Classes}(R)$, $\mathrm{rng}(t) \cap E \neq \emptyset$. So, let $E \in \mathit{Classes}(R)$ be given, and let $p \in E$ be arbitrary. Then, clearly, $(t \circ t')(p) \in \mathrm{rng}(t) \cap E$.

(b) $\Rightarrow$ (c): Immediate.

(c) $\Rightarrow$ (a): Suppose (c). Further suppose that $\theta$ is an eps, and that $t : \mathbb{N} \to \mathbb{N}$ witnesses $\theta \leq \psi$. Then, by (c), there exists a ceer $R \subseteq \mathrm{Equiv}(\psi)$ such that, for all but finitely many $E \in \mathit{Classes}(R)$, $\mathrm{rng}(t) \cap E \neq \emptyset$. Let $n$ be the number of elements of $\mathit{Classes}(R)$ that do *not* intersect $\mathrm{rng}(t)$, and let $E_0, ..., E_{n-1}$ be those elements. Choose $q_0, ..., q_{n-1}$ such that, for each $i < n$ and $p \in E_i$, $\theta_{q_i} = \psi_p$. Note that, for each $p$, either $R$ equates $p$ to some element of $\mathrm{rng}(t)$, or $p \in E_i$, for some $i < n$. It follows that the function $t' : \mathbb{N} \to \mathbb{N}$, defined next, is computable.

$$t' = \lambda p \,\textbf{.}\, \begin{cases} q, \text{ where } q \text{ is first found such that } \langle p, t(q)\rangle \in R, \\ \quad \text{if such a } q \text{ exists;} \\ q_i, \text{ otherwise, where } i \text{ is such that } p \in E_i. \end{cases} \tag{9}$$

It is straightforward to verify that $t'$ witnesses $\psi \leq \theta$. $\square$ (**Theorem 6**)

Theorem 7, restated just below, is our second main result. It establishes that there there exists a minimal eps $\psi$ such that, for each ceer $R$, either $R$ contradicts $\psi$'s program equivalence relation, or there exists a translation function $t$ into $\psi$ such that the range of $t$ *fails* to intersect *infinitely many* of $R$'s equivalence classes.

**Theorem 7.** There exists an eps $\psi$ satisfying (a) and (b) below.

(a) $\psi$ is minimal.
(b) For each ceer $R$, there exists a translation function $t$ into $\psi$ such that $R$ does *not* weakly tie $t$ into $\psi$.

The proof of Theorem 7 makes use of the following lemma.

**Lemma 10.** Let $J_0, ..., J_{n-1}$ be any finite collection of computably enumerable sets. Then, there exists an infinite, computable set $X$, and a finite set $L \subseteq \{0, ..., n-1\}$, such that, for each $x \in X$ and $\ell < n$, $x \in J_\ell$ iff $\ell \in L$.

*Proof.* Let $J_0, ..., J_{n-1}$ be as stated. The set $X$ is the set $X_n$, constructed as follows. Set $X_0 = \mathbb{N}$. Then, for each $\ell < n$, act according to the following conditions.

- COND. (a) $[J_\ell \cap X_\ell$ is infinite$]$. Set $X_{\ell+1}$ to any infinite, computable subset of $J_\ell \cap X_\ell$.
- COND. (b) $[J_\ell \cap X_\ell$ is finite$]$. Set $X_{\ell+1} = \{x \in X_\ell \mid x > \max(J_\ell \cap X_\ell)\}$.

The set $L$ is such that

$$L = \{\ell \mid \text{cond. (a) applies for } \ell\}. \tag{10}$$

Clearly, $X$ is infinite and computable. Further note that

$$X_0 \supseteq X_1 \supseteq \cdots \supseteq X_n. \tag{11}$$

It is easily seen that, for each $\ell < n$: if $\ell \in L$, then $J_\ell \supseteq X_{\ell+1}$; whereas, if $\ell \notin L$, then $J_\ell \cap X_{\ell+1} = \emptyset$. It then follows from (11) that, for each $x \in X_n$ and $\ell < n$, $x \in J_\ell$ iff $\ell \in L$. $\square$ (**Lemma 10**)

*Proof of Theorem 7.* The eps $\psi$ is constructed below, following some necessary definitions. Let $\mathit{Aux} \subseteq \mathit{PartComp}$ be such that

$$\mathit{Aux} = \mathit{PartComp} \setminus \{\langle i, j\rangle^{<k+1} \mid i, j \in \mathbb{N} \wedge k < 2^i\}. \tag{12}$$

It is straightforward to show that $\mathit{Aux}$ is 1-1, computably enumerable. So, let $(\alpha_\ell)_{\ell \in \mathbb{N}}$ be a 1-1, effective numbering of $\mathit{Aux}$.

As is common, $\psi$ is constructed in stages, i.e., $\psi$ is the union of $\psi^0 \subseteq \psi^1 \subseteq \cdots$. In conjunction with $\psi$, four computable predicates are constructed: $\lambda i, s \,\textbf{.}\, [i \in R\text{-flags}^s]$, $\lambda i, j, \ell, s \,\textbf{.}\, [\langle i, j, \ell\rangle \in t\text{-flags}^s]$, $\lambda \ell, s \,\textbf{.}\, [\ell \in \mathrm{Src}^s]$, and $\lambda p, s \,\textbf{.}\, [p \in \mathrm{Dst}^s]$. The purposes of these predicates are as follows.

- The $R$-flags predicate keeps track of which $i$ are such that $W_i$ contradicts $\psi$'s program equivalence relation. More precisely, for each $i$, if there exists an $s$ such that $i \in R\text{-flags}^s$, then $W_i \not\subseteq \text{Equiv}(\psi)$.
- The $t$-flags predicate helps to keep track of which $\ell$ *may be* such that $\varphi_\ell$ is a translation function into $\psi$. It will turn out that: if $i$ and $\ell$ are such that $W_i \subseteq \text{Equiv}(\psi)$ and $\varphi_\ell$ is a translation function into $\psi$, then, for each $j$, and all but finitely many $s$, $\langle i, j, \ell \rangle \in t\text{-flags}^s$.
- The Src predicate keeps track of which $\ell$ are such that $\alpha_\ell$ has not yet been assigned to any $\psi$-program. In particular, if $\ell$ and $s$ are such that $\ell \in \text{Src}^s$ and $\alpha_\ell \neq \lambda x.\uparrow$, then, for each $p$, $\psi_p^s \neq \alpha_\ell$.
- The Dst predicate keeps track of which $\psi$-programs have not yet been used. More precisely, if $p$ and $s$ are such that $p \in \text{Dst}^s$, then $\psi_p^s = \lambda x.\uparrow$.

For each $i$ and $s$, $i \in R\text{-flags}^{s+1}$ iff $i \in R\text{-flags}^s$, unless stated otherwise. Analogous statements apply to the $t$-flags, Src, and Dst predicates, as well. The following will be clear from the construction of $\psi$, for each $s$.

$$R\text{-flags}^s \subseteq R\text{-flags}^{s+1}. \tag{13}$$
$$t\text{-flags}^s \subseteq t\text{-flags}^{s+1}. \tag{14}$$
$$\text{Src}^s \supseteq \text{Src}^{s+1}. \tag{15}$$
$$\text{Dst}^s \supseteq \text{Dst}^{s+1}. \tag{16}$$

Let height : $\mathbb{N}^3 \to \mathbb{N}$ be such that, for each $i$, $j$, and $s$,

$$\text{height}_{i,j}^s = |\{\ell \mid \langle i, j, \ell \rangle \in t\text{-flags}^s\}|. \tag{17}$$

It will be clear from the construction of $\psi$ that, for each $i$, $j$, $\ell$, and $s$,

$$\langle i, j, \ell \rangle \in t\text{-flags}^s \Rightarrow \ell < i. \tag{18}$$

Thus, for each $i$, $j$, and $s$,

$$\text{height}_{i,j}^s \leq i. \tag{19}$$

Let num : $\mathbb{N}^3 \to \mathbb{N}$ be such that, for each $i$, $j$, and $s$,

$$\text{num}_{i,j}^s = 2^{i-h}, \text{ where } h = \text{height}_{i,j}^s. \tag{20}$$

Let $f : \mathbb{N}^3 \to \mathbb{N}$ be such that, for each $i$, $j$, and $k$,

$$f_{i,j}(k) = 2\langle i, j \cdot 2^{i+1} + k \rangle. \tag{21}$$

For each $i$, $j$, $s$, and $k < \text{num}_{i,j}^s$, let $E_{i,j,k}^s \in \mathcal{F}\!in$ and $\bar{E}_{i,j,k}^s \in \mathcal{F}\!in$ be as follows, with $h = \text{height}_{i,j}^s$.

$$E_{i,j,k}^s = \{f_{i,j}(k \cdot 2^{h+1}), ..., f_{i,j}(k \cdot 2^{h+1} + 2^h - 1)\}. \tag{22}$$
$$\bar{E}_{i,j,k}^s = \{f_{i,j}(k \cdot 2^{h+1} + 2^h), ..., f_{i,j}((k+1) \cdot 2^{h+1} - 1)\}. \tag{23}$$

Note that, for each $i$, $j$, and $s$, if one lets $h = \text{height}_{i,j}^s$, and it happens that $\text{height}_{i,j}^{s+1} = h + 1$, then, for each $k < \text{num}_{i,j}^{s+1}$,

$$\begin{aligned}
E_{i,j,k}^{s+1} &= \{f_{i,j}(k \cdot 2^{h+2}), ..., f_{i,j}(k \cdot 2^{h+2} + 2^{h+1} - 1)\} \\
&= \{f_{i,j}(k \cdot 2^{h+2}), ..., f_{i,j}(k \cdot 2^{h+2} + 2^h - 1)\} \\
&\quad \cup \{f_{i,j}(k \cdot 2^{h+2} + 2^h), ..., f_{i,j}(k \cdot 2^{h+2} + 2^{h+1} - 1)\} \\
&= \{f_{i,j}(2k \cdot 2^{h+1}), ..., f_{i,j}(2k \cdot 2^{h+1} + 2^h - 1)\} \\
&\quad \cup \{f_{i,j}(2k \cdot 2^{h+1} + 2^h), ..., f_{i,j}(2k \cdot 2^{h+1} + 2^{h+1} - 1)\} \\
&= \{f_{i,j}(2k \cdot 2^{h+1}), ..., f_{i,j}(2k \cdot 2^{h+1} + 2^h - 1)\} \\
&\quad \cup \{f_{i,j}(2k \cdot 2^{h+1} + 2^h), ..., f_{i,j}((2k+1) \cdot 2^{h+1} - 1)\} \\
&= E_{i,j,2k}^s \cup \bar{E}_{i,j,2k}^s.
\end{aligned} \tag{24}$$

It can be shown that, under the same conditions,

$$\bar{E}_{i,j,k}^{s+1} = E_{i,j,2k+1}^s \cup \bar{E}_{i,j,2k+1}^s. \tag{25}$$

- STAGE $s = -1$. Do the following.
  - Set $R\text{-flags}^0 = \emptyset$.
  - Set $t\text{-flags}^0 = \emptyset$.
  - Set $\text{Src}^0 = \mathbb{N}$.
  - Set $\text{Dst}^0 = 2\mathbb{N} + 1$.
  - For each $i$, $j$, and $k < 2^i$, set $\psi^0_{f_{i,j}(2k)} = \psi^0_{f_{i,j}(2k+1)} = \langle i, j \rangle^{<k+1}$.
  - For each $p \in 2\mathbb{N} + 1$, set $\psi^0_p = \lambda x.\uparrow$.
- STAGE $s = \langle 0, \ell \rangle$. If $\ell \in \text{Src}^s$, then do the following.
  - Set $\text{Src}^{s+1} = \text{Src}^s \setminus \{\ell\}$.
  - Set $\text{Dst}^{s+1} = \text{Dst}^s \setminus \{\min \text{Dst}^s\}$.
  - Set $\psi^{s+1}_{\min \text{Dst}^s} = \alpha_\ell$.
- STAGE $s = \langle i+1, 0, - \rangle$. Determine whether there exist $j$ and $k$ satisfying conditions (a)-(c) just below.
  (a) $i \notin R\text{-flags}^s$.
  (b) $k < \text{num}^s_{i,j}$.
  (c) $W^s_i \cap (E^s_{i,j,k} \times \bar{E}^s_{i,j,k}) \neq \emptyset$.
  If such $j$ and $k$ exist, then do the following.
  - Set $R\text{-flags}^{s+1} = R\text{-flags}^s \cup \{i\}$.
  - Choose any $\ell, m \in \text{Src}^s$ such that $\ell \neq m$ and $\langle i, j \rangle^{<2^i} \subseteq \alpha_\ell \cap \alpha_m$.
  - Let $d : \mathbb{N} \to \mathbb{N}$ be any 1-1, computable function such that $\text{rng}(d)$ is computable, $\text{rng}(d) \subseteq \text{Dst}^s$, and $\text{Dst}^s \setminus \text{rng}(d)$ is infinite.
  - Set $\text{Src}^{s+1} = \text{Src}^s \setminus \{\ell, m\}$.
  - Set $\text{Dst}^{s+1} = \text{Dst}^s \setminus \text{rng}(d)$.
  - For each $j$, each $k < \text{num}^s_{i,j}$, and each $p \in E^s_{i,j,k}$, set $\psi^{s+1}_p = \alpha_\ell$.
  - For each $j$, each $k < \text{num}^s_{i,j}$, and each $q \in \bar{E}^s_{i,j,k}$, set $\psi^{s+1}_q = \alpha_m$.
  - For each $j$ and $k < \text{num}^s_{i,j}$, set $\psi^{s+1}_{d(n+k)} = \langle i, j \rangle^{<(k+1) \cdot 2^h}$, where $n = \sum_{\hat{j} < j} \text{num}^s_{i,\hat{j}}$ and $h = \text{height}^s_{i,j}$.
- STAGE $s = \langle i+1, j+1, \ell, - \rangle$. Let $h = \text{height}^s_{i,j}$. Determine whether conditions (i)-(iv) just below are satisfied.
  (i) $\ell < i$.
  (ii) $i \notin R\text{-flags}^s$.
  (iii) $\langle i, j, \ell \rangle \notin t\text{-flags}^s$.
  (iv) For each $k < \text{num}^s_{i,j}$, $\text{rng}(\varphi^s_\ell) \cap (E^s_{i,j,k} \cup \bar{E}^s_{i,j,k}) \neq \emptyset$.
  If so, then do the following.
  - Set $t\text{-flags}^{s+1} = t\text{-flags}^s \cup \{\langle i, j, \ell \rangle\}$. (Note that this implies $\text{height}^{s+1}_{i,j} = \text{height}^s_{i,j} + 1$.)
  - Let $n = \text{num}^{s+1}_{i,j}$. (Note that, by the just previous step, $n = \text{num}^s_{i,j}/2$.)
  - Let $\{q_0 < q_1 < \cdots < q_{n-1}\}$ be the $n$ least elements of $\text{Dst}^s$.
  - Set $\text{Dst}^{s+1} = \text{Dst}^s \setminus \{q_0, q_1, ..., q_{n-1}\}$.
  - For each $k < n$ and $p \in (E^{s+1}_{i,j,k} \cup \bar{E}^{s+1}_{i,j,k})$, set $\psi^{s+1}_p = \langle i, j \rangle^{(2k+2) \cdot 2^h}$.
  - For each $k < n$, set $\psi^{s+1}_{q_k} = \langle i, j \rangle^{<(2k+1) \cdot 2^h}$.

**Fig. 2.** The construction of $\psi$ in the proof of Theorem 7. The symbols height, num, $f$, $E$, and $\bar{E}$ are defined in (17), (20), (21), (22), and (23), respectively.

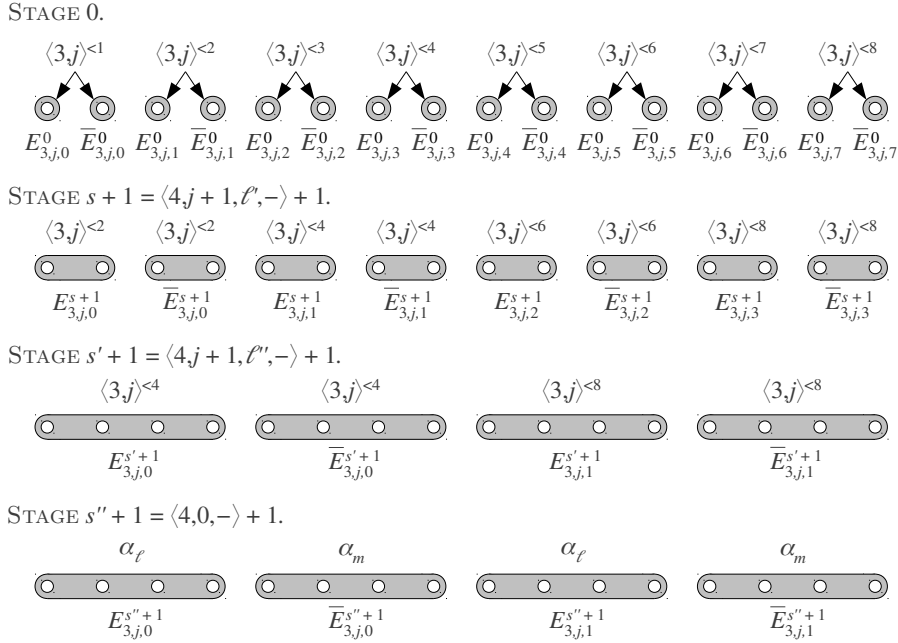

**Fig. 3.** A depiction of what *could* happen in the proof of Theorem 7 with respect to the $\psi$-programs of the form $f_{3,j}(k)$, where $j$ is arbitrary and $k < 16$ (see text).

The partial function $\psi$ is constructed in Figure 2. To help to give some of the intuition behind the construction, Figure 3 depicts what *could* happen with respect to the $\psi$-programs of the form $f_{3,j}(k)$, where $j$ is arbitrary and $k < 16$. In stage 0, the programs will form *eight* pairs of equivalence classes, where the $k$th pair computes $\langle 3, j \rangle^{<k+1}$ (the first such pair being the 0th). If, subsequently, the conditions of some stage $s$ of the form $\langle 4, j+1, \ell', - \rangle$ are satisfied, then, in stage $s+1$, the programs will form *four* pairs of equivalence classes, where the $k$th pair computes $\langle 3, j \rangle^{<2k+2}$. If, similarly, the conditions of some stage $s'$ of the form $\langle 4, j+1, \ell'', - \rangle$ are satisfied (where $\ell' \neq \ell''$), then, in stage $s'+1$, the programs will form *two* pairs of equivalence classes, where the $k$th pair computes $\langle 3, j \rangle^{<4k+4}$. If, finally, the conditions of some stage $s''$ of the form $\langle 4, 0, - \rangle$ are satisfied, then, in stage $s''+1$, the equivalence classes will alternate in computing $\alpha_\ell$ and $\alpha_m$, for some *distinct* $\ell$ and $m$.

Note that by (14), (17), and (19), the following function $\text{height}^\infty : \mathbb{N}^2 \to \mathbb{N}$ is well-defined. For each $i$ and $j$,
$$\text{height}_{i,j}^\infty = \max\{\text{height}_{i,j}^s \mid s \in \mathbb{N}\}. \tag{26}$$
For each $i$ and $j$, let $\text{num}_{i,j}^\infty$ be defined in a manner analogous to (20), but with $h = \text{height}_{i,j}^\infty$. For each $i$, $j$, and $k < \text{num}_{i,j}^\infty$, let $E_{i,j,k}^\infty$ and $\bar{E}_{i,j,k}^\infty$ be defined in a manner analogous to (22) and (23) (respectively), but with $h = \text{height}_{i,j}^\infty$.

Claim 7.1 below establishes that $\psi$ is an eps. Claim 7.7 below establishes that $\psi$ satisfies (a) in the statement of the theorem, i.e., that $\psi$ is minimal. Claim 7.8 below establishes that $\psi$ satisfies (b) in the statement of the theorem, i.e., that for each ceer $R$, there exists a translation function $t$ into $\psi$ such that $R$ does *not* weakly tie $t$ into $\psi$.

**Claim 7.1.** $\psi$ is an eps.

*Proof of Claim.* Clearly, $\psi$ is partial computable. Thus, it suffices to show that, for each $\zeta \in \text{PartComp}$, there exists a $p$ such that $\psi_p = \zeta$. So, let $\zeta \in \text{PartComp}$ be given. Consider the following cases.

CASE [$\zeta \in \text{Aux}$]. Let $\ell$ be such that $\alpha_\ell = \zeta$, and let $s = \langle 0, \ell \rangle$. Then, the following are easily verifiable from the construction of $\psi$.

– If $\ell \notin \text{Src}^s$, then there exists a $p$ of the form $f_{i,j}(k)$, for some $i$, $j$, and $k$, such that $\psi_p^s = \zeta$.

- If $\ell \in \text{Src}^s$, then $\psi^{s+1}_{\min \text{Dst}^s} = \zeta$.

CASE $[\zeta \notin \mathit{Aux}]$. Let $i$, $j$, $k$, and $h$ be such that $\zeta = \langle i,j \rangle^{<(2k+1)\cdot 2^h}$. Then, the following are easily verifiable from the construction of $\psi$.

- If $\text{height}^\infty_{i,j} \leq h$ and $(\forall s)[i \notin R\text{-flags}^s]$, then, for each $p$,
$$p \in \{f_{i,j}((2k+1)\cdot 2^{h+1} - 2), f_{i,j}((2k+1)\cdot 2^{h+1} - 1)\} \;\Rightarrow\; \psi_p = \zeta. \tag{27}$$

- If $\text{height}^\infty_{i,j} > h$ or $(\exists s)[i \in R\text{-flags}^s]$, then there exists a $p \in \text{Dst}^0 \;(= 2\mathbb{N}+1)$ such that $\psi_p = \zeta$.

$\hfill \square$ (**Claim 7.1**)

**Claim 7.2.** Suppose that $i$ is such that $(\forall s)[i \notin R\text{-flags}^s]$. Then, for each $j$, each $k < \text{num}^\infty_{i,j}$, and each $p$,
$$p \in (E^\infty_{i,j,k} \cup \bar{E}^\infty_{i,j,k}) \;\Leftrightarrow\; \psi_p = \langle i,j \rangle^{(k+1)\cdot 2^h}, \tag{28}$$
where $h = \text{height}^\infty_{i,j}$.

*Proof of Claim.* Easily verifiable from the construction of $\psi$. $\hfill \square$ (**Claim 7.2**)

**Claim 7.3.** Suppose that $i$ is such that $(\exists s)[i \in R\text{-flags}^s]$. Let $s_{\min}$ be *least* such that $i \in R\text{-flags}^{s_{\min}+1}$. Then, there exist *distinct* $\ell$ and $m$ such that (a) and (b) below.

(a) For each $p$,
$$p \in \bigcup \{E^{s_{\min}}_{i,j,k} \mid j \in \mathbb{N} \;\wedge\; k < \text{num}^{s_{\min}}_{i,j}\} \;\Leftrightarrow\; \psi_p = \alpha_\ell. \tag{29}$$

(b) For each $q$,
$$q \in \bigcup \{\bar{E}^{s_{\min}}_{i,j,k} \mid j \in \mathbb{N} \;\wedge\; k < \text{num}^{s_{\min}}_{i,j}\} \;\Leftrightarrow\; \psi_q = \alpha_m. \tag{30}$$

*Proof of Claim.* Easily verifiable from the construction of $\psi$. $\hfill \square$ (**Claim 7.3**)

**Claim 7.4.** For each $p \in \text{Dst}^0 \;(= 2\mathbb{N}+1)$ and $q$, if $\psi_p = \psi_q$, then $p = q$.

*Proof of Claim.* Easily verifiable from the construction of $\psi$. $\hfill \square$ (**Claim 7.4**)

**Claim 7.5.** Suppose that $i$, $j$, $\ell$, and $s$ are such that $\langle i,j,\ell \rangle \in t\text{-flags}^s$. Then,
$$\text{rng}(\varphi_\ell) \cap E^s_{i,j,k} \neq \emptyset \;\wedge\; \text{rng}(\varphi_\ell) \cap \bar{E}^s_{i,j,k} \neq \emptyset. \tag{31}$$

*Proof of Claim.* Suppose that $i$, $j$, $\ell$, and $s$ are as stated. Let $s_{\min}$ be *least* such that
$$\langle i,j,\ell \rangle \in t\text{-flags}^{s_{\min}+1}. \tag{32}$$

Thus, $s > s_{\min}$. By the construction of $\psi$, for each $k' < \text{num}^{s_{\min}}_{i,j}$,
$$\text{rng}(\varphi_\ell) \cap (E^{s_{\min}}_{i,j,k'} \cup \bar{E}^{s_{\min}}_{i,j,k'}) \neq \emptyset. \tag{33}$$

It follows from (24) and (32) that, for each $s > s_{\min}$ and $k < \text{num}^s_{i,j}$, there exists a $k' < \text{num}^{s_{\min}}_{i,j}$ such that
$$E^{s_{\min}}_{i,j,k'} \cup \bar{E}^{s_{\min}}_{i,j,k'} \subseteq E^s_{i,j,k}. \tag{34}$$

Similarly, it follows from (25) and (32) that, for each $s > s_{\min}$ and $k < \text{num}^s_{i,j}$, there exists a $k' < \text{num}^{s_{\min}}_{i,j}$ such that
$$E^{s_{\min}}_{i,j,k'} \cup \bar{E}^{s_{\min}}_{i,j,k'} \subseteq \bar{E}^s_{i,j,k}. \tag{35}$$

Formula (31) is implied by (33), (34), and (35). $\hfill \square$ (**Claim 7.5**)

For each $i$, $j$, and $s$, act according to the following computable conditions. (Note that cond. (a) is computable, in part, because there are only finitely many $i \leq \ell$.)

- COND. (a) $\left[\text{height}^s_{i,j} < \text{height}^{s+1}_{i,j} \wedge i \leq \ell \wedge (\forall s)[i \notin R\text{-flags}^s]\right]$. For each $k < \text{num}^{s+1}_{i,j}$ and
$$p, q \in (E^{s+1}_{i,j,k} \cup \bar{E}^{s+1}_{i,j,k}),$$
list $\langle p, q \rangle$ into $R$.
- COND. (b) $\left[\text{height}^s_{i,j} < \text{height}^{s+1}_{i,j} \wedge i > \ell\right]$. For each $k < \text{num}^{s+1}_{i,j}$ and
$$p, q \in E^{s+1}_{i,j,k},$$
list $\langle p, q \rangle$ into $R$. Similarly, for each
$$p, q \in \bar{E}^{s+1}_{i,j,k},$$
list $\langle p, q \rangle$ into $R$.

For each $i$, act according to the following partial computable condition.

- COND. (c) $(\exists s)[i \in R\text{-flags}^s]$. Let $s_{\min}$ be *least* such that $i \in R\text{-flags}^{s_{\min}+1}$, and do the following. For each
$$p, q \in \bigcup \{E^{s_{\min}}_{i,j,k} \mid j \in \mathbb{N} \wedge k < \text{num}^{s_{\min}}_{i,j}\},$$
list $\langle p, q \rangle$ into $R$. Similarly, for each
$$p, q \in \bigcup \{\bar{E}^{s_{\min}}_{i,j,k} \mid j \in \mathbb{N} \wedge k < \text{num}^{s_{\min}}_{i,j}\},$$
list $\langle p, q \rangle$ into $R$.

**Fig. 4.** The construction of $R$ in the proof of Claim 7.7.

**Claim 7.6.** Suppose that $i$ is such that $W_i \subseteq \text{Equiv}(\psi)$. Then, $(\forall s)[i \notin R\text{-flags}^s]$.

*Proof of Claim.* The proof is by contrapositive. Suppose that $i$ is such that $(\exists s)[i \in R\text{-flags}^s]$. Let $s_{\min}$ be *least* such that $i \in R\text{-flags}^{s_{\min}+1}$. Then, by the construction of $\psi$, there exist $j$ and $k$ such that
$$W^{s_{\min}}_i \cap (E^{s_{\min}}_{i,j,k} \times \bar{E}^{s_{\min}}_{i,j,k}) \neq \emptyset. \tag{36}$$

Furthermore, by Claim 7.3($\Rightarrow$), there exist *distinct* $\ell$ and $m$ such that (a) and (b) below.

(a) For each $p \in E^{s_{\min}}_{i,j,k}$, $\psi_p = \alpha_\ell$.
(b) For each $q \in \bar{E}^{s_{\min}}_{i,j,k}$, $\psi_q = \alpha_m$.

Since $\alpha$ is 1-1 and $\ell \neq m$, $\alpha_\ell \neq \alpha_m$. Thus, by (36) and (a) and (b) just above, $W_i \not\subseteq \text{Equiv}(\psi)$. □ (**Claim 7.6**)

**Claim 7.7.** $\psi$ satisfies (a) in the statement of the theorem, i.e., $\psi$ is minimal.

*Proof of Claim.* Let $t$ be any translation function into $\psi$, and let $\ell$ be such that $\varphi_\ell = t$. To show the claim, a ceer $R$ is exhibited such that $R$ strongly ties $t$ into $\psi$. Initially, $R$ consists of $\{\langle p, p \rangle \mid p \in \mathbb{N}\}$. Then, pairs are added to $R$ as in Figure 4.

Clearly, $R$ is a ceer. That $R \subseteq \text{Equiv}(\psi)$ follows from the ($\Rightarrow$) directions of Claims 7.2 and 7.3. It remains to show that, for each $E \in \textit{Classes}(R)$, $\text{rng}(t) \cap E \neq \emptyset$. It is straightforward to verify that each $E \in \textit{Classes}(R)$ is of one of the following four types.

- TYPE I. $E$ is of the form
$$E^\infty_{i,j,k} \cup \bar{E}^\infty_{i,j,k}, \tag{37}$$
where: $i \leq \ell$, $(\forall s)[i \notin R\text{-flags}^s]$, $j$ is arbitrary, and $k < \text{num}^\infty_{i,j}$. (Intuitively, $E$ is the result of one or more invocations of cond. (a) in Figure 4.)

– TYPE II. Either $E$ is of the form
$$E^{\infty}_{i,j,k} \tag{38}$$
or $E$ is of the form
$$\bar{E}^{\infty}_{i,j,k} \tag{39}$$
where: $i > \ell$, $(\forall s)[i \notin R\text{-flags}^s]$, $j$ is arbitrary, and $k < \text{num}^{\infty}_{i,j}$. (Intuitively, $E$ is the result of one or more invocations of cond. (b) in Figure 4.)

– TYPE III. Either $E$ is of the form
$$\bigcup \{E^{s_{\min}}_{i,j,k} \mid j \in \mathbb{N} \wedge k < \text{num}^{s_{\min}}_{i,j}\} \tag{40}$$
or $E$ is of the form
$$\bigcup \{\bar{E}^{s_{\min}}_{i,j,k} \mid j \in \mathbb{N} \wedge k < \text{num}^{s_{\min}}_{i,j}\} \tag{41}$$
where: $i$ is such that $(\exists s)[i \in R\text{-flags}^s]$, and $s_{\min}$ is *least* such that $i \in R\text{-flags}^{s_{\min}+1}$. (Intuitively, $E$ is the result of zero or more invocations of cond. (b) in Figure 4, followed by a single invocation of cond. (c).)

– TYPE IV. $E = \{p\}$, for some $p \in \text{Dst}^0$ $(= 2\mathbb{N} + 1)$.

Let $E \in \text{Classes}(R)$ be given. If $E$ is of type I, then it follows from Claim 7.2($\Leftarrow$) that $\text{rng}(t) \cap E \neq \emptyset$. If $E$ is of type III, then it follows from Claim 7.3($\Leftarrow$) that $\text{rng}(t) \cap E \neq \emptyset$. If $E$ is of type IV, then it follows from Claim 7.4 that $\text{rng}(t) \cap E \neq \emptyset$.

So, suppose that $E$ is of type II. Let $i$, $j$, and $k$ be such that $E = E^{\infty}_{i,j,k}$ or $\bar{E} = \bar{E}^{\infty}_{i,j,k}$, as appropriate. Further suppose, by way of contradiction, that $\text{rng}(t) \cap E = \emptyset$. Thus,
$$\text{rng}(t) \cap E^{\infty}_{i,j,k} = \emptyset \ \vee\ \text{rng}(t) \cap \bar{E}^{\infty}_{i,j,k} = \emptyset. \tag{42}$$

Note that by Claim 7.2($\Leftarrow$), for each $k' < \text{num}^{\infty}_{i,j}$ and $p$,
$$\psi_p = \langle i,j \rangle^{(k'+1)\cdot 2^h} \Rightarrow p \in (E^{\infty}_{i,j,k'} \cup \bar{E}^{\infty}_{i,j,k'}), \tag{43}$$
where $h = \text{height}^{\infty}_{i,j}$. Thus, since $t$ is a translation function into $\psi$, it must be the case that, for each $k' < \text{num}^{\infty}_{i,j}$,
$$\text{rng}(t) \cap (E^{\infty}_{i,j,k'} \cup \bar{E}^{\infty}_{i,j,k'}) \neq \emptyset. \tag{44}$$
Choose $s$ such that $s$ is of the form $\langle i+1, j+1, \ell, -\rangle$, $\text{height}^s_{i,j} = \text{height}^{\infty}_{i,j}$, and, for each $k' < \text{num}^{\infty}_{i,j}$,
$$\text{rng}(\varphi^s_\ell) \cap (E^{\infty}_{i,j,k'} \cup \bar{E}^{\infty}_{i,j,k'}) \neq \emptyset. \tag{45}$$
Note that by (42) and Claim 7.5, $\langle i,j,\ell \rangle \notin t\text{-flags}^s$. It follows that all of the conditions of stage $s$ are satisfied. Thus, $\langle i,j,\ell \rangle \in t\text{-flags}^{s+1}$. But then
$$\text{height}^{s+1}_{i,j} > \text{height}^s_{i,j} = \text{height}^{\infty}_{i,j} \tag{46}$$
— a contradiction. $\square$ (**Claim 7.7**)

**Claim 7.8.** $\psi$ satisfies (b) in the statement of the theorem, i.e., for each ceer $R$, there exists a translation function $t$ into $\psi$ such that $R$ does *not* weakly tie $t$ into $\psi$.

*Proof of Claim.* Suppose that ceer $R$ is such that
$$R \subseteq \text{Equiv}(\psi). \tag{47}$$
Let $i$ be such that $W_i = R$. Note that by Claim 7.6,
$$(\forall s)[i \notin R\text{-flags}^s]. \tag{48}$$
For each $\ell < i$, let $J_\ell$ be as follows.
$$J_\ell = \{j \mid (\exists s)[\langle i,j,\ell \rangle \in t\text{-flags}^s]\}. \tag{49}$$

Clearly, for each $\ell < i$, $J_\ell$ is computably enumerable. Thus, by Lemma 10, there exists an infinite, computable set $X$, and a finite set $L \subseteq \{0, ..., i-1\}$, such that, for each $x \in X$ and $\ell \in L$, $x \in J_\ell$ iff $\ell \in L$. Thus, for each $x \in X$,

$$\begin{aligned} L &= \{\ell \mid x \in J_\ell\} \\ &= \{\ell \mid x \in \{j \mid (\exists s)[\langle i, j, \ell \rangle \in t\text{-flags}^s]\}\} \\ &= \{\ell \mid (\exists s)[\langle i, x, \ell \rangle \in t\text{-flags}^s]\}. \end{aligned}$$

It follows that, for each $x \in X$, $\text{height}_{i,x}^\infty = |L|$ and $\text{num}_{i,j}^\infty = 2^{i-|L|}$. Let $t$ be any computable function such that

$$\text{rng}(t) = \mathbb{N} \setminus \bigcup \{E_{i,x,0}^\infty \mid x \in X\}.^4 \tag{50}$$

It is straightforward to show that that $t$ is a translation function into $\psi$. On the other hand, it is clearly the case that, for each $x \in X$,

$$\text{rng}(t) \cap E_{i,x,0}^\infty = \emptyset. \tag{51}$$

Thus, to complete the proof, it suffices to show that, for each $E \in \text{Equiv}(R)$ and $x \in X$,

$$E \cap E_{i,x,0}^\infty \neq \emptyset \Rightarrow E \subseteq E_{i,x,0}^\infty. \tag{52}$$

By way of contradiction, suppose otherwise, as witnessed by $E$ and $x$, i.e.,

$$E \cap E_{i,x,0}^\infty \neq \emptyset \wedge E \nsubseteq E_{i,x,0}^\infty. \tag{53}$$

By (47), (48), (53), and Claim 7.2 (both directions), it must be the case that

$$E \cap \bar{E}_{i,x,0}^\infty \neq \emptyset., \tag{54}$$

Thus, by the first conjunct of (53), and by (54), there exists a stage $s$ of the form $\langle i+1, 0, - \rangle$ in which all of the conditions of that stage are satisfied. Thus, $i \in R\text{-flags}^{s+1}$. But this contradicts (48). $\square$ (**Claim 7.8**)

$\square$ (**Theorem 7**)

Theorem 8, restated just below, is our third main result. It establishes that the strong and weak notions of Definition 5 separate when one considers single equivalence relations.

**Theorem 8.** There exists an eps $\psi$ and a ceer $R \subseteq \text{Equiv}(\psi)$ satisfying (a) and (b) below.

(a) For each translation function $t$ into $\psi$, $R$ weakly ties $t$ into $\psi$.
(b) For each ceer $R'$, there exists a translation function $t$ into $\psi$ such that $R'$ does *not* strongly tie $t$ into $\psi$.

*Proof.* The proof is essentially a modification to the proof of Theorem 7. Intuitively, one eliminates all uses of $j$ in that proof. So, for example, for each $i$, rather than start with infinitely many pairs of equivalence classes,

$$\{(E_{i,j,k}^0, \bar{E}_{i,j,k}^0) \mid j \in \mathbb{N} \wedge k < 2^i\}, \tag{55}$$

one instead starts with just $2^i$ many such pairs,

$$\{(E_{i,k}^0, \bar{E}_{i,k}^0) \mid k < 2^i\}. \tag{56}$$

This has the effect of invalidating Claim 7.8 (and of making Lemma 10 unnecessary).

Let $\mathit{Aux} \subseteq \mathit{PartComp}$ be such that

$$\mathit{Aux} = \mathit{PartComp} \setminus \{i^{<k+1} \mid i \in \mathbb{N} \wedge k < 2^i\}. \tag{57}$$

Let $(\alpha_\ell)_{\ell \in \mathbb{N}}$ be a 1-1, effective numbering of $\mathit{Aux}$.

In conjunction with $\psi$, four computable predicates are constructed: $\lambda i, s.[i \in R\text{-flags}^s]$, $\lambda i, \ell, s.[\langle i, \ell \rangle \in t\text{-flags}^s]$, $\lambda \ell, s.[\ell \in \text{Src}^s]$, and $\lambda p, s.[p \in \text{Dst}^s]$. The purposes of these predicates are similar to those in the proof of Theorem 7. (Note, however, the difference in the type of the $t$-flags predicate.)

Let $f : \mathbb{N}^2 \to \mathbb{N}$ be such that, for each $i$ and $k$,

$$f_i(k) = 2 \cdot (2^{i+1} + k - 2). \tag{58}$$

The following symbols are defined in a manner analogous to the proof of Theorem 7.

---

[4] In (50), we chose to use $\bigcup\{E_{i,x,0}^\infty \mid x \in X\}$. But the proof can be completed using $\bigcup\{E_{i,x,k}^\infty \mid x \in X\}$ or $\bigcup\{\bar{E}_{i,x,k}^\infty \mid x \in X\}$, for any $k < \min\{\text{num}_{i,x}^\infty \mid x \in X\}$.

- height : $\mathbb{N}^2 \to \mathbb{N}$ and $\text{height}^\infty : \mathbb{N} \to \mathbb{N}$
- num : $\mathbb{N}^2 \to \mathbb{N}$ and $\text{num}^\infty : \mathbb{N} \to \mathbb{N}$

The following symbols are defined similarly, but with $f$ as in (58).

- $E : \mathbb{N}^3 \to \mathit{Fin}$ and $E^\infty : \mathbb{N}^2 \to \mathit{Fin}$
- $\bar{E} : \mathbb{N}^3 \to \mathit{Fin}$ and $\bar{E}^\infty : \mathbb{N}^2 \to \mathit{Fin}$

Suppose that $i$ and $s$ are such that $\text{height}_i^{s+1} = \text{height}_i^s + 1$. Then, by reasoning in a manner analogous to (24), it can be shown that, for each $k < \text{num}_i^{s+1}$, the following.

$$\begin{aligned} E_{i,k}^{s+1} &= E_{i,2k}^s \cup \bar{E}_{i,2k}^s \\ \bar{E}_{i,k}^{s+1} &= E_{i,2k+1}^s \cup \bar{E}_{i,2k+1}^s. \end{aligned} \tag{59}$$

The partial function $\psi$ is constructed in Figure 5. One can show Claims 8.1 through 8.6 below. The proofs are similar to those of Claims 7.1 through 7.6 (respectively).

**Claim 8.1.** $\psi$ is an eps.

**Claim 8.2.** Suppose that $i$ is such that $(\forall s)[i \notin R\text{-flags}^s]$. Then, for each $k < \text{num}_i^\infty$, and each $p$,

$$p \in (E_{i,k}^\infty \cup \bar{E}_{i,k}^\infty) \Leftrightarrow \psi_p = i^{(k+1) \cdot 2^h}, \tag{60}$$

where $h = \text{height}_i^\infty$.

**Claim 8.3.** Suppose that $i$ is such that $(\exists s)[i \in R\text{-flags}^s]$. Let $s_{\min}$ be *least* such that $i \in R\text{-flags}^{s_{\min}+1}$. Then, there exist *distinct* $\ell$ and $m$ such that (a) and (b) below.

(a) For each $p$,

$$p \in \bigcup \{E_{i,k}^{s_{\min}} \mid k < \text{num}_i^{s_{\min}}\} \Leftrightarrow \psi_p = \alpha_\ell. \tag{61}$$

(b) For each $q$,

$$q \in \bigcup \{\bar{E}_{i,k}^{s_{\min}} \mid k < \text{num}_i^{s_{\min}}\} \Leftrightarrow \psi_q = \alpha_m. \tag{62}$$

**Claim 8.4.** For each $p \in \text{Dst}^0 \ (= 2\mathbb{N} + 1)$ and $q$, if $\psi_p = \psi_q$, then $p = q$.

**Claim 8.5.** Suppose that $i$, $\ell$, and $s$ are such that $\langle i, \ell \rangle \in t\text{-flags}^s$. Then,

$$\text{rng}(\varphi_\ell) \cap E_{i,k}^s \neq \emptyset \ \land \ \text{rng}(\varphi_\ell) \cap \bar{E}_{i,k}^s \neq \emptyset. \tag{63}$$

**Claim 8.6.** Suppose that $i$ is such that $W_i \subseteq \text{Equiv}(\psi)$. Then, $(\forall s)[i \notin R\text{-flags}^s]$.

The relation $R$ consists initially of $\{\langle p, p \rangle \mid p \in \mathbb{N}\}$. Then, pairs are added to $R$ as in Figure 6.

Clearly, $R$ is a ceer. That $R \subseteq \text{Equiv}(\psi)$ follows from the ($\Rightarrow$) directions of Claims 8.2 and 8.3.

Claim 8.7 below establishes that $\psi$ and $R$ satisfy (a) in the statement of the theorem, i.e., that for each translation function $t$ into $\psi$, $R$ weakly ties $t$ into $\psi$. Claim 8.8 below establishes that $\psi$ satisfies (b) in the statement of the theorem, i.e., that for each ceer $R'$, there exists a translation function $t$ into $\psi$ such that $R'$ does *not* strongly tie $t$ into $\psi$.

- STAGE $s = -1$. Do the following.
  - Set $R\text{-flags}^0 = \emptyset$.
  - Set $t\text{-flags}^0 = \emptyset$.
  - Set $\text{Src}^0 = \mathbb{N}$.
  - Set $\text{Dst}^0 = 2\mathbb{N} + 1$.
  - For each $i$ and $k < 2^i$, set $\psi^0_{f_i(2k)} = \psi^0_{f_i(2k+1)} = i^{<k+1}$.
  - For each $p \in 2\mathbb{N} + 1$, set $\psi^0_p = \lambda x.\uparrow$.
- STAGE $s = \langle 0, \ell \rangle$. If $\ell \in \text{Src}^s$, then do the following.
  - Set $\text{Src}^{s+1} = \text{Src}^s \setminus \{\ell\}$.
  - Set $\text{Dst}^{s+1} = \text{Dst}^s \setminus \{\min \text{Dst}^s\}$.
  - Set $\psi^{s+1}_{\min \text{Dst}^s} = \alpha_\ell$.
- STAGE $s = \langle i+1, 0, - \rangle$. Determine whether there exists a $k$ satisfying conditions (a)-(c) just below.
  (a) $i \notin R\text{-flags}^s$.
  (b) $k < \text{num}^s_i$.
  (c) $W^s_i \cap (E^s_{i,k} \times \bar{E}^s_{i,k}) \neq \emptyset$.
  If such a $k$ exists, then do the following.
  - Set $R\text{-flags}^{s+1} = R\text{-flags}^s \cup \{i\}$.
  - Choose any $\ell, m \in \text{Src}^s$ such that $\ell \neq m$ and $i^{<2^i} \subseteq \alpha_\ell \cap \alpha_m$.
  - Let $n = \text{num}^s_i$.
  - Let $\{p_0 < p_1 < \cdots < p_{n-1}\}$ be the $n$ least elements of $\text{Dst}^s$.
  - Set $\text{Src}^{s+1} = \text{Src}^s \setminus \{\ell, m\}$.
  - Set $\text{Dst}^{s+1} = \text{Dst}^s \setminus \{p_0, p_1, ..., p_{n-1}\}$.
  - For each $k < n$ and $p \in E^s_{i,k}$, set $\psi^{s+1}_p = \alpha_\ell$.
  - For each $k < n$ and $q \in \bar{E}^s_{i,k}$, set $\psi^{s+1}_q = \alpha_m$.
  - For each $k < n$, set $\psi^{s+1}_{p_k} = i^{<(k+1) \cdot 2^h}$, where $h = \text{height}^s_i$.
- STAGE $s = \langle i+1, \ell+1, - \rangle$. Let $h = \text{height}^s_i$. Determine whether conditions (i)-(iv) just below are satisfied.
  (i) $\ell < i$.
  (ii) $i \notin R\text{-flags}^s$.
  (iii) $\langle i, \ell \rangle \notin t\text{-flags}^s$.
  (iv) For each $k < \text{num}^s_i$, $\text{rng}(\varphi^s_\ell) \cap (E^s_{i,k} \cup \bar{E}^s_{i,k}) \neq \emptyset$.
  If so, then do the following.
  - Set $t\text{-flags}^{s+1} = t\text{-flags}^s \cup \{\langle i, \ell \rangle\}$. (Note that this implies $\text{height}^{s+1}_i = \text{height}^s_i + 1$.)
  - Let $n = \text{num}^{s+1}_i$. (Note that, by the just previous step, $n = \text{num}^s_i/2$.)
  - Let $\{q_0 < q_1 < \cdots < q_{n-1}\}$ be the $n$ least elements of $\text{Dst}^s$.
  - Set $\text{Dst}^{s+1} = \text{Dst}^s \setminus \{q_0, q_1, ..., q_{n-1}\}$.
  - For each $k < n$ and $p \in (E^{s+1}_{i,k} \cup \bar{E}^{s+1}_{i,k})$, set $\psi^{s+1}_p = i^{(2k+2) \cdot 2^h}$.
  - For each $k < n$, set $\psi^{s+1}_{q_k} = i^{<(2k+1) \cdot 2^h}$.

**Fig. 5.** The construction of $\psi$ in the proof of Theorem 8.

For each $i$ and $s$, act according to the following computable condition.

- COND. (a) [$\text{height}_i^s < \text{height}_i^{s+1}$]. For each $k < \text{num}_i^{s+1}$ and
$$p, q \in E_{i,k}^{s+1},$$
list $\langle p, q \rangle$ into $R$. Similarly, for each
$$p, q \in \bar{E}_{i,k}^{s+1},$$
list $\langle p, q \rangle$ into $R$.

For each $i$, act according to the following partial computable condition.

- COND. (b) $(\exists s)[i \in R\text{-flags}^s]$. Let $s_{\min}$ be *least* such that $i \in R\text{-flags}^{s_{\min}+1}$, and do the following. For each
$$p, q \in \bigcup \{E_{i,k}^{s_{\min}} \mid k < \text{num}_i^{s_{\min}}\},$$
list $\langle p, q \rangle$ into $R$. Similarly, for each
$$p, q \in \bigcup \{\bar{E}_{i,k}^{s_{\min}} \mid k < \text{num}_i^{s_{\min}}\},$$
list $\langle p, q \rangle$ into $R$.

**Fig. 6.** The construction of $R$ in the proof of Theorem 8.

**Claim 8.7.** $\psi$ and $R$ satisfy (a) in the statement of the theorem, i.e., for each translation function $t$ into $\psi$, $R$ weakly ties $t$ into $\psi$.

*Proof of Claim.* It is straightforward to verify that each $E \in \textit{Classes}(R)$ is of one of the following three types.

- TYPE I. Either $E$ is of the form
$$E_{i,k}^\infty \tag{64}$$
or $E$ is of the form
$$\bar{E}_{i,k}^\infty \tag{65}$$
where: $i$ is such that $(\forall s)[i \notin R\text{-flags}^s]$, and $k < \text{num}_i^\infty$. (Intuitively, $E$ is the result of one or more invocations of cond. (a) in Figure 6.)
- TYPE II. Either $E$ is of the form
$$\bigcup \{E_{i,k}^{s_{\min}} \mid k < \text{num}_i^{s_{\min}}\} \tag{66}$$
or $E$ is of the form
$$\bigcup \{\bar{E}_{i,k}^{s_{\min}} \mid k < \text{num}_i^{s_{\min}}\} \tag{67}$$
where: $i$ is such that $(\exists s)[i \in R\text{-flags}^s]$, and $s_{\min}$ is *least* such that $i \in R\text{-flags}^{s_{\min}+1}$. (Intuitively, $E$ is the result of zero or more invocations of cond. (a) in Figure 6, followed by a single invocation of cond. (b).)
- TYPE III. $E = \{p\}$, for some $p \in \text{Dst}^0 \ (= 2\mathbb{N} + 1)$.

Let $t$ be any translation function into $\psi$, and let $\ell$ be such that $\varphi_\ell = t$. Note that there are only finitely many $E \in \textit{Classes}(R)$ of type I for which $i \leq \ell$, where $i$ is such that $E = E_{i,k}^\infty$ or $E = \bar{E}_{i,k}^\infty$, as appropriate. Thus, to show the claim, it suffices to show that, for each $E \in \textit{Classes}(R)$: if $E$ is of type II or III, then $\text{rng}(t) \cap E \neq \emptyset$; whereas, if $E$ is of type I, then $\text{rng}(t) \cap E \neq \emptyset$ or $i \leq \ell$ (where $i$ is as just mentioned).

So, let $E \in \textit{Classes}(R)$ be given. If $E$ is of type II, then it follows from Claim 8.3($\Leftarrow$) that $\text{rng}(t) \cap E \neq \emptyset$. If $E$ is of type III, then it follows from Claim 8.4 that $\text{rng}(t) \cap E \neq \emptyset$.

So, suppose that $E$ is of type I, and that $\text{rng}(t) \cap E = \emptyset$. Let $i$ and $k$ be such $E = E_{i,k}^\infty$ or $E = \bar{E}_{i,k}^\infty$, as appropriate. To show that $i \leq \ell$, one first assumes otherwise, by way of contradiction. One then proceeds in a manner analogous to the proof of Claim 7.7, beginning just before (42). □ (**Claim 8.7**)

**Claim 8.8.** $\psi$ satisfies (b) in the statement of the theorem, i.e., for each ceer $R'$, there exists a translation function $t$ into $\psi$ such that $R'$ does *not* strongly tie $t$ into $\psi$.

*Proof of Claim.* Suppose that ceer $R' \subseteq \text{Equiv}(\psi)$. Let $i$ be such that $W_i = R'$. Let $t$ be any computable function such that
$$\text{rng}(t) = \mathbb{N} \setminus E_{i,0}^{\infty}. \tag{68}$$
It is straightforward to show that $t$ is a translation function into $\psi$. On the other hand, it is clearly the case that
$$\text{rng}(t) \cap E_{i,0}^{\infty} = \emptyset. \tag{69}$$
Thus, to complete the proof, it suffices to show that, for each $E \in \text{Equiv}(R)$,
$$E \cap E_{i,0}^{\infty} \neq \emptyset \;\Rightarrow\; E \subseteq E_{i,0}^{\infty}. \tag{70}$$
This can be shown in a manner analogous to the proof of Claim 7.8, beginning just after (52).

$\hspace{10cm}\square$ (**Claim 8.8**)

$\hspace{10cm}\square$ (**Theorem 8**)

Theorem 9, restated just below, is our final main result. It establishes that there can exist a single ceer that strongly ties each translation function into an eps, yet that eps's program equivalence relation can fail to be computably enumerable.

**Theorem 9.** *There exists an eps $\psi$ and a ceer $R \subseteq \text{Equiv}(\psi)$ satisfying (a) and (b) below.*

(a) *For each translation function $t$ into $\psi$, $R$ strongly ties $t$ into $\psi$.*
(b) *$\text{Equiv}(\psi)$ is not computably enumerable.*

*Proof.* The eps $\psi$ is constructed below, following some necessary definitions. Let $\mathcal{A}ux \subseteq \mathcal{P}art\mathcal{C}omp$ be such that
$$\mathcal{A}ux = \mathcal{P}art\mathcal{C}omp \setminus (\{i^{<j+1} \mid i, j \in \mathbb{N}\} \cup \{\lambda x.i \mid i \in \mathbb{N}\}). \tag{71}$$
Let $(\alpha_k)_{k \in \mathbb{N}}$ be a 1-1, effective numbering of $\mathcal{A}ux$.

In conjunction with $\psi$, the following six computable predicates are constructed.

- $\lambda i, s.[i \in R\text{-flags}^s]$
- $\lambda i, j, s.[\langle i, j \rangle \in t\text{-flags}^s]$
- $\lambda j, s.[j \in \text{Src}^s]$
- $\lambda p, s.[p \in \text{Dst}^s]$
- $\lambda p, i, s.[p \in E_i^s]$
- $\lambda q, i, s.[q \in \bar{E}_i^s]$

The purposes of these predicates are similar to those in the proofs of Theorems 7 and 8. Note, however, that in the proofs of Theorems 7 and 8, the $E$ and $\bar{E}$ predicates were *calculated*; whereas, in this proof, they are *constructed*. The following will be clear from the construction of $\psi$, for each $i$ and $s$.
$$E_i^s \subseteq E_i^{s+1}. \tag{72}$$
$$\bar{E}_i^s \subseteq \bar{E}_i^{s+1}. \tag{73}$$
For each $i$, let $E_i^{\infty}$ and $\bar{E}_i^{\infty}$ be as follows.
$$E_i^{\infty} = \bigcup \{E_i^s \mid s \in \mathbb{N}\}. \tag{74}$$
$$\bar{E}_i^{\infty} = \bigcup \{\bar{E}_i^s \mid s \in \mathbb{N}\}. \tag{75}$$

The partial function $\psi$ is constructed in Figure 7. Claim 9.1 below establishes that $\psi$ is an eps.

**Claim 9.1.** $\psi$ *is an eps.*

*Proof of Claim.* Clearly, $\psi$ is partial computable. Thus, it suffices to show that, for each $\zeta \in \mathcal{P}art\mathcal{C}omp$, there exists a $p$ such that $\psi_p = \zeta$. So, let $\zeta \in \mathcal{P}art\mathcal{C}omp$ be given. Consider the following cases.

CASE $[\zeta \in \mathcal{A}ux]$. Let $k$ be such that $\alpha_k = \zeta$, and let $s = \langle 0, k \rangle$. Then, the following are easily verifiable from the construction of $\psi$.

- STAGE $s = -1$. Do the following.
  - Set $R\text{-flags}^0 = \emptyset$.
  - Set $t\text{-flags}^0 = \emptyset$.
  - Set $\text{Src}^0 = \mathbb{N}$.
  - Set $\text{Dst}^0 = 3\mathbb{N} + 2$.
  - For each $i$, set $E_i^0 = \bar{E}_i^0 = \emptyset$.
  - For each $i$ and $j$, set $\psi_{3\langle i,j\rangle}^0 = i^{<2j+1}$.
  - For each $i$ and $j$, set $\psi_{3\langle i,j\rangle+1}^0 = i^{<2j+2}$.
  - For each $p \in 3\mathbb{N} + 2$, set $\psi_p^0 = \lambda x.\uparrow$.
- STAGE $s = \langle 0, k\rangle$. If $k \in \text{Src}^s$, then do the following.
  - Set $\text{Src}^{s+1} = \text{Src}^s \setminus \{k\}$.
  - Set $\text{Dst}^{s+1} = \text{Dst}^s \setminus \{\min \text{Dst}^s\}$.
  - Set $\psi_{\min \text{Dst}^s}^{s+1} = \alpha_k$.
- STAGE $s = \langle i+1, 0, -\rangle$. If $i \notin R\text{-flags}^s$ and $W_i^s \cap (E_i^s \times \bar{E}_i^s) \neq \emptyset$, then do the following.
  - Set $R\text{-flags}^{s+1} = R\text{-flags}^s \cup \{i\}$.
  - Let $n$ be *least* such that, for each $p \in (E_i^s \cup \bar{E}_i^s)$, $\psi_p \subseteq i^{<n}$.
  - Choose any $k, \ell \in \text{Src}^s$ such that $k \neq \ell$ and $i^{<n} \subseteq \alpha_k \cap \alpha_\ell$.
  - Set $\text{Src}^{s+1} = \text{Src}^s \setminus \{k, \ell\}$.
  - Set $\text{Dst}^{s+1} = \text{Dst}^s \setminus \{\min \text{Dst}^s\}$.
  - For each $p \in E_i^s$, set $\psi_p^{s+1} = \alpha_k$.
  - For each $q \in \bar{E}_i^s$, set $\psi_q^{s+1} = \alpha_\ell$.
  - Set $\psi_{\min \text{Dst}^s}^{s+1} = \lambda x.i$.
- STAGE $s = \langle i+1, j+1, -\rangle$. Determine whether conditions (i)-(iii) just below are satisfied.
  (i) $i \notin R\text{-flags}^s$.
  (ii) $\langle i, j\rangle \notin t\text{-flags}^s$.
  (iii) $\{3\langle i,j\rangle, 3\langle i,j\rangle + 1\} \subseteq \text{rng}(\varphi_j^s)$.
  If so, then do the following.
  - Set $t\text{-flags}^{s+1} = t\text{-flags}^s \cup \{\langle i,j\rangle\}$.
  - Set $E_i^{s+1} = E_i^s \cup \{3\langle i,j\rangle\}$.
  - Set $\bar{E}_i^{s+1} = \bar{E}_i^s \cup \{3\langle i,j\rangle + 1\}$.
  - Let $n$ be *least* such that, for each $p \in (E_i^{s+1} \cup \bar{E}_i^{s+1})$, $\psi_p^{s+1} \subseteq i^{<n}$.
  - For each $p \in (E_i^{s+1} \cup \bar{E}_i^{s+1})$, set $\psi_p = i^{<n}$.
  - Let $\{q_0 < q_1\}$ be the two *least* elements of $\text{Dst}^s$.
  - Set $\text{Dst}^{s+1} = \text{Dst}^s \setminus \{q_0, q_1\}$.
  - Set $\psi_{q_0}^{s+1} = i^{<2j+1}$.
  - Set $\psi_{q_1}^{s+1} = i^{<2j+2}$.

**Fig. 7.** The construction of $\psi$ in the proof of Theorem 9.

- If $k \notin \mathrm{Src}^s$, then there exists a $p$ of the form $3\langle i,j\rangle$ or $3\langle i,j\rangle + 1$, for some $i$ and $j$, such that $\psi_p^s = \zeta$.
- If $k \in \mathrm{Src}^s$, then $\psi_{\min \mathrm{Dst}^s}^{s+1} = \zeta$.

CASE $\bigl[\zeta \notin \mathit{Aux} \wedge (\exists i,j)[\zeta = i^{<2j+1}]\bigr]$. Let $i$ and $j$ be as in the case. Then, the following are easily verifiable from the construction of $\psi$.

- If $(\forall s)[\langle i,j\rangle \notin t\text{-flags}^s]$, then $\psi_{3\langle i,j\rangle} = \zeta$.
- If $(\exists s)[\langle i,j\rangle \in t\text{-flags}^s]$, then there exists a $p \in \mathrm{Dst}^0\ (= 3\mathbb{N}+2)$ such that $\psi_p = \zeta$.

CASE $\bigl[\zeta \notin \mathit{Aux} \wedge (\exists i,j)[\zeta = i^{<2j+2}]\bigr]$. Similar to the previous case. $\square$ (**Claim 9.1**)

The relation $R$ is defined as follows.

$$R = \begin{aligned}&\{\langle p,p\rangle \mid p \in \mathbb{N}\}\\ &\cup\, \{\langle p,q\rangle \mid p,q \in E_i^\infty \wedge i \in \mathbb{N}\}\\ &\cup\, \{\langle p,q\rangle \mid p,q \in \bar{E}_i^\infty \wedge i \in \mathbb{N}\}.\end{aligned} \tag{76}$$

Clearly, $R$ is a ceer. That $R \subseteq \mathrm{Equiv}(\psi)$ follows from the $(\Rightarrow)$ directions of Claims 9.2 and 9.3.

Claim 9.6 below establishes that $\psi$ and $R$ satisfy (a) in the statement of the theorem, i.e., that for each translation function $t$ into $\psi$, $R$ strongly ties $t$ into $\psi$. Claim 9.7 below establishes that $\psi$ satisfies (b) in the statement of the theorem, i.e., that $\mathrm{Equiv}(\psi)$ is *not* computably enumerable.

**Claim 9.2.** Suppose that $i$ is such that $(\forall s)[i \notin R\text{-flags}^s]$. Then, (a) and (b) below.

(a) Each of $E_i^\infty$ and $\bar{E}_i^\infty$ is infinite.
(b) For each $p$, $p \in (E_i^\infty \cup \bar{E}_i^\infty) \Leftrightarrow \psi_p = \lambda x\,.\,i$.

*Proof of Claim.* Suppose that $i$ is such that $(\forall s)[i \notin R\text{-flags}^s]$. Note that there exist infinitely many $j$ such that $\mathrm{rng}(\varphi_j) = \mathbb{N}$. Thus, there exist infinitely many $j$ such that $\{3\langle i,j\rangle, 3\langle i,j\rangle + 1\} \subseteq \mathrm{rng}(\varphi_j^s)$, for all but finitely many $s$. It follows that there exist infinitely many stages of the form $\langle i+1, j+1, -\rangle$ such that all of the conditions of those stages are satisfied. Given this fact, both (a) and (b) are easily verifiable from the construction of $\psi$. $\square$ (**Claim 9.2**)

**Claim 9.3.** Suppose that $i$ is such that $(\exists s)[i \in R\text{-flags}^s]$. Let $s_{\min}$ be *least* such that $i \in R\text{-flags}^{s_{\min}+1}$. Then, (a) and (b) below.

(a) $E_i^\infty = E_i^{s_{\min}}$ and $\bar{E}_i^\infty = \bar{E}_i^{s_{\min}}$.
(b) There exist *distinct* $k$ and $\ell$ such that (i) and (ii) below.
 (i) For each $p$, $p \in E_i^\infty \Leftrightarrow \psi_p = \alpha_k$.
 (ii) For each $q$, $q \in \bar{E}_i^\infty \Leftrightarrow \psi_q = \alpha_\ell$.

*Proof of Claim.* Easily verifiable from the construction of $\psi$. $\square$ (**Claim 9.3**)

**Claim 9.4.** For each $p$ such that

$$p \notin \bigcup \{E_i^\infty \cup \bar{E}_i^\infty \mid i \in \mathbb{N}\}, \tag{77}$$

and, for each $q$, if $\psi_p = \psi_q$, then $p = q$.

*Proof of Claim.* Easily verifiable from the construction of $\psi$. $\square$ (**Claim 9.4**)

**Claim 9.5.** Suppose that $i$ is such that $W_i \subseteq \mathrm{Equiv}(\psi)$. Then, $(\forall s)[i \notin R\text{-flags}^s]$.

*Proof of Claim.* The proof is by contrapositive. Suppose that $i$ is such that $(\exists s)[i \in R\text{-flags}^s]$. Let $s_{\min}$ be *least* such that $i \in R\text{-flags}^{s_{\min}+1}$. Then, by the construction of $\psi$,

$$W_i^{s_{\min}} \cap (E_i^{s_{\min}} \times \bar{E}_i^{s_{\min}}) \neq \emptyset. \tag{78}$$

Furthermore, by Claim 9.3$(\Rightarrow)$, there exist *distinct* $k$ and $\ell$ such that (a) and (b) below.

(a) For each $p \in E_i^{s_{\min}}$, $\psi_p = \alpha_k$.
(b) For each $q \in \bar{E}_i^{s_{\min}}$, $\psi_q = \alpha_\ell$.

Since $\alpha$ is 1-1 and $k \neq \ell$, $\alpha_k \neq \alpha_\ell$. Thus, by (78) and (a) and (b) just above, $W_i \not\subseteq \mathrm{Equiv}(\psi)$. $\square$ (**Claim 9.5**)

**Claim 9.6.** $\psi$ and $R$ satisfy (a) in the statement of the theorem, i.e., for each translation function $t$ into $\psi$, $R$ strongly ties $t$ into $\psi$.

*Proof of Claim.* It is straightforward to verify that each $E \in \mathit{Classes}(R)$ is of one of the following three types.

- TYPE I. Either $E$ is of the form $E_i^\infty$ or $E$ is of the form $\bar{E}_i^\infty$ where: $i$ is such that $(\forall s)[i \notin R\text{-flags}^s]$.
- TYPE II. Either $E$ is of the form $E_i^\infty$ or $E$ is of the form $\bar{E}_i^\infty$ where: $i$ is such that $(\exists s)[i \in R\text{-flags}^s]$.
- TYPE III. $E = \{p\}$, for some $p$ such that

$$p \notin \bigcup \{E_i^\infty \cup \bar{E}_i^\infty \mid i \in \mathbb{N}\}. \tag{79}$$

Let $t$ be any translation function into $\psi$, and let $E \in \mathit{Classes}(R)$ be given. If $E$ is of type I, then it follows from Claim 9.2($\Leftarrow$) that $\mathrm{rng}(t) \cap E \neq \emptyset$. If $E$ is of type II, then it follows from Claim 9.3($\Leftarrow$) that $\mathrm{rng}(t) \cap E \neq \emptyset$. If $E$ is of type III, then it follows from Claim 9.4 that $\mathrm{rng}(t) \cap E \neq \emptyset$. $\square$ (**Claim 9.6**)

**Claim 9.7.** $\psi$ satisfies (b) in the statement of the theorem, i.e., $\mathrm{Equiv}(\psi)$ is *not* computably enumerable.

*Proof of Claim.* By way of contradiction, let $i$ be such that

$$W_i = \mathrm{Equiv}(\psi). \tag{80}$$

By (80) and Claim 9.5,

$$(\forall s)[i \notin R\text{-flags}^s]. \tag{81}$$

By (81) and Claim 9.2(a), each of $E_i^\infty$ and $\bar{E}_i^\infty$ is infinite, and, thus,

$$\text{each of } E_i^\infty \text{ and } \bar{E}_i^\infty \text{ is } \textit{non}\text{-empty.} \tag{82}$$

By (81) and Claim 9.2(b)($\Rightarrow$), for each $p \in (E_i^\infty \cup \bar{E}_i^\infty)$,

$$\psi_p = \lambda x \,.\, i. \tag{83}$$

By (80), (82), and (83),

$$W_i \cap (E_i^\infty \times \bar{E}_i^\infty) \neq \emptyset. \tag{84}$$

By (81) and (84), there exists a stage $s$ of the form $\langle i+1, 0, -\rangle$ in which all of the conditions of that stage are satisfied. But then $i \in R\text{-flags}^{s+1}$, contradicting (81). $\square$ (**Claim 9.7**)

$\square$ (**Theorem 9**)

## References

Ers68. Y. L. Ershov. On computable enumerations. *Algebra and Logic*, 7(5):330–346, 1968.

FKW82. R. Freivalds, E. B. Kinber, and R. Wiehagen. Inductive inference and computable one-one numberings. *Zeitschrift für Mathematische Logik und Grundlagen der Mathematik*, 28(27-32):463–479, 1982.

Fri58. R. M. Friedberg. Three theorems on recursive enumeration. I. Decomposition. II. Maximal Set. III. Enumeration without duplication. *Journal of Symbolic Logic*, 23(3):309–316, 1958.

GG01. S. Gao and P. Gerdes. Computably enumerable equivalence relations. *Studia Logica*, 67(1):27–59, 2001.

GYY93. S. Goncharov, A. Yakhnis, and V. Yakhnis. Some effectively infinite classes of enumerations. *Annals of Pure and Applied Logic*, 60(3):207–235, 1993.

HK94. E. Herrmann and M. Kummer. Diagonals and D-maximal sets. *Journal of Symbolic Logic*, 59(1):60–72, 1994.

JST11. S. Jain, F. Stephan, and J. Teutsch. Index sets and universal numberings. *Journal of Computer and System Sciences*, 77(4):760–773, 2011.

Khu69a. A. B. Khutoretskii. On the reducibility of computable numerations. *Algebra and Logic*, 8(2):145–151, 1969.

Khu69b. A. B. Khutoretskii. Two existence theorems for computable numerations. *Algebra and Logic*, 8(4):277–282, 1969.

Kum89. M. Kummer. A note on direct sums of Friedbergnumberings. *Journal of Symbolic Logic*, 54(3):1009–1010, 1989.

Kum90. M. Kummer. An easy priority-free proof of a theorem of Friedberg. *Theoretical Computer Science*, 74(2):249–251, 1990.

Lav77. I. A. Lavrov. Computable numberings. In *Logic, Foundations of Mathematics and Computability Theory*, pages 195–206, 1977.

Mal65. A. I. Mal'cev. Positive and negative numerations. *Proceedings of the USSR Academy of Sciences*, 160(2):278–280, 1965.

Mal71. A. I. Mal'cev. Positive and negative numberings. In *The Metamathematics of Algebraic Systems*, volume 66 of *Studies in Logic and the Foundations of Mathematics*, pages 379–383. Elsevier, 1971. Translated by B. F. Wells III.

Moe12. S. E. Moelius III. Characteristics of minimal effective programming systems. In *Proceedings of Computability in Europe 2012 (CiE 2012) - How the World Computes*, Lecture Notes in Computer Science. Springer, 2012. To appear.

MWY78. M. Machtey, K. Winklmann, and P. Young. Simple Gödel numberings, isomorphisms, and programming properties. *SIAM Journal on Computing*, 7(1):39–60, 1978.

PE64. M. B. Pour-El. Gödel numberings versus Friedberg numberings. *Proceedings of the American Mathematical Society*, 15(2):252–256, 1964.

Ric81. G. A. Riccardi. The independence of control structures in abstract programming systems. *Journal of Computer and System Sciences*, 22(2):107–143, 1981.

Rog58. H. Rogers, Jr. Gödel numberings of partial recursive functions. *Journal of Symbolic Logic*, 23(3):331–341, 1958.

Rog67. H. Rogers, Jr. *Theory of Recursive Functions and Effective Computability*. McGraw Hill, 1967. Reprinted, MIT Press, 1987.

Roy87. J. S. Royer. *A Connotational Theory of Program Structure*, volume 273 of *Lecture Notes in Computer Science*. Springer, 1987.

Sch82. B. Schinzel. On decomposition of Gödelnumberings into Friedbergnumberings. *Journal of Symbolic Logic*, 47(2):267–274, 1982.

Spr90. D. Spreen. Computable one-to-one enumerations of effective domains. *Information and Computation*, 84(1):26–46, 1990.